\documentclass[bibnotes,a4paper,twocolumn,prb]{revtex4}%
\usepackage{amsfonts}
\usepackage{amsmath}
\usepackage{amssymb}
\usepackage{graphicx}%
\graphicspath{{Z:/SUPRA/fluctuations_fflo/article_PRB/}}
\ifx\pdfoutput\relax\let\pdfoutput=\undefined\fi
\newcount\msipdfoutput
\ifx\pdfoutput\undefined\else
\ifcase\pdfoutput\else
\ifx\paperwidth\undefined\else
\ifdim\paperheight=0pt\relax\else\pdfpageheight\paperheight\fi
\ifdim\paperwidth=0pt\relax\else\pdfpagewidth\paperwidth\fi
\fi\fi\fi
\begin{document}
\title[Fluctuations near to an iso/anisotropic FFLO state]{Oscillations of magnetization and conductivity in anisotropic
Fulde-Ferrell-Larkin-Ovchinnikov superconductors}
\author{Fran\c{c}ois K\textsc{onschelle}}
\author{J\'{e}r\^{o}me C\textsc{ayssol} }
\author{Alexandre I. B\textsc{uzdin}}
\altaffiliation{Also at \textit{Institut Universitaire de France}}

\email{a.bouzdine@cpmoh.u-bordeaux1.fr}
\affiliation{Universit\'{e} de Bordeaux ; CNRS ; CPMOH, F-33405 Talence, France}
\date{\today}
\startpage{1}


\begin{abstract}
We derive the fluctuational magnetization and the paraconductivity of
Fulde-Ferrell-Larkin-Ovchinnikov (FFLO) superconductors in their normal state.
The FFLO superconducting fluctuations induce oscillations of the magnetization
between diamagnetism and unusual paramagnetism which originates from the
competition between paramagnetic and orbital effects. We also predict a strong
anisotropy of the paraconductivity when the FFLO transition is approached in
contrast with the case of a uniform BCS state. Finally building a
Ginzburg-Levanyuk argument, we demonstrate that these fluctuation effects can
be safely treated within the Gaussian approximation since the critical
fluctuations are proeminent only within an experimentally inaccessible
temperature interval.

\end{abstract}
\maketitle

\section{Introduction.}

Forty years ago, Fulde and Ferrell \cite{fulde_ferrell(1964)}, and Larkin and
Ovchinnikov \cite{larkin_ovchinnikov(1964)} predicted that the paramagnetism
of the electron gas might induce a novel superconducting state wherein the
order parameter is modulated in real-space. In their original proposal, these
authors considered a singlet $s$-wave superconductor perturbed by the Zeeman
effect (paramagnetic effect), and neglected completely the orbital coupling
and the disorder. For most type-II superconductors, the superconductivity is
destroyed by the orbital pair-breaking effect which leads to a more familiar
inhomogeneous superconducting state: the Abrikosov vortex lattice. In order to
observe the Fulde-Ferrell-Larkin-Ovchinnikov (FFLO) state, the paramagnetic
effect must break Cooper pairs more efficiently than the orbital one. Such a
situation may be realized in tridimensional (3D) superconductors with large
internal exchange fields, like the rare-earth magnetic superconductor
ErRh$_{4}$B$_{4}$, see \cite{bulaevskii85} for a review. Another possibility
corresponds to a quasi two-dimensional (2D) layered superconductor wherein the
weakness of the interplane hopping suppresses the orbital effect for in-plane
magnetic field. Being the ratio of the critical fields, $H_{c}^{orb}(T=0)$ in
the pure orbital limit, and $H_{p}(T=0)$ in the pure paramagnetic limit, the
Maki parameter $\alpha_{m}=\sqrt{2}H_{c}^{orb}(T=0)/H_{p}(T=0)$ quantifies the
relative strength of those pair breaking mechanisms. Besides demanding a large
Maki parameter ($\alpha_{m}>1.8$), the occurance of the FFLO state also
requires very clean samples since it is far less robust against disorder than
the usual vortex lattice, see \cite{casalbuoni.nardulli.2004} and
\cite{shimahara_matsuda} for recent reviews.

Recently, there have been mounting evidence that the heavy fermion
superconductor CeCoIn$_{5}$ under magnetic field might fullfil those
stringents conditions \cite{radovan03,bianchi02,bianchi03}, although the
magnetism of this system is still under debate. A superconducting phase have
been reported at large magnetic field and low temperature which is distinct
from the uniform superconducting phase realized at lower fields. The
characteristics of this phase depends upon the orientation of the field
relatively to the basal plane of the tetragonal CeCoIn$_{5}$ lattice
\cite{shimahara_matsuda}. In the field-induced organic
superconductor\cite{uji.2006} $\lambda$-(BETS)$_{2}$FeCl$_{4}$, and in the
layered organic superconductor\cite{lortz.2007} $\kappa$-(BEDT-TTF)$_{2}%
$Cu(NCS)$_{2}$ the FFLO state have been reported when a strong magnetic field
(20 T for latter one) is applied along the superconducting planes.

However in practice, the identification of the FFLO state is hindered by the
interplay between orbital and paramagnetic effects. The first available
experimental clue is the shape of the transition line $H_{c}(T)$ separating
the normal state from the inhomogeneous superconducting state. A lot of
theoretical works have been devoted to the description of this $H_{c}(T)$
line. For moderate Maki parameters, $\alpha_{m}<9$, the structure of the FFLO
modulation involves a zero Landau level (index $n=0$) function (Gaussian with
no additional modulation) \cite{gruenberg66}. For higher Maki parameter,
$\alpha_{m}>9$, the Cooper pair wave function of a 3D superconductor consists
in a cascade of more exotic solutions, the so-called multi-quanta states,
which are described by a higher (index $n>0$) Landau level \cite{buzdin963D}.
Such values of Maki parameters are rather high for 3D compounds (for instance
CeCoIn$_{5}$ has $\alpha_{m}=4.6-5$) but they can be achieved in layered
quasi-2D superconductors (or superconducting thin films) under in-plane
magnetic fields \cite{buzdin962D}. All these studies were performed so far in
the framework of isotropic models, namely for the idealistic case of a
spherical Fermi surface in the normal state. Moreover it has been shown that
an elliptic Fermi surface leads to the same phenomenology at cost of
introducing an angle-dependent Maki parameter \cite{buzdinelliptic}.

In real compounds, the crystal lattice (or the pairing symmetry) induces a non
trivial anisotropy which matters a lot for the modulated state
\cite{yang_sondhi.1998,vorontsov_sauls_graf.2005} since it essentially lifts
the degeneracy between various orientations of the FFLO modulation. Recently,
the interplay of paramagnetic and orbital effects was reconsidered in the
presence of such a non trivial anisotropy, namely for a Fermi surface which
slightly differs from the spherical or elliptical shape \cite{denisov09}.
Using a perturbative approach, it was found that even a small anisotropy
stabilizes the exotic multi-quanta states which can therefore exist at lower
Maki parameter (any $\alpha_{m}>1.8$) than predicted by the idealized
isotropic models. According to this prediction such states are likely to occur
in any real anisotropic Pauli limited superconductor. More specifically in the
tetragonal symmetry, 3 scenarios are possible for the FFLO state: a) Maximal
FFLO modulation along the field with zero Landau level state, b) Highest
Landau level modulation in the plane perpendicular to the field and no FFLO
modulation, and c) Both Landau level and FFLO modulations. This three
scenarios correspond to the tetragonal symmetry and was derived within a
single Landau level approximation, which is valid at large field. It may thus
be relevant to explain the observation of two high-field and low temperature
phases of CeCoIn$_{5}$, which exhibit contrasted behaviors under distinct
magnetic field orientations (inside or perpendicular to the CeIn$_{3}$
planes). Nevertheless the shape of the $H_{c}(T)$ transition line is far from
sufficient to establish a clear correspondence between one phase and a
particular class of solutions among the three a)-c) possibilities. It is thus
necessary to gain complementary informations to determine which scenario among
a)-c) is actually realized. As natural precussors of the transition, the
fluctuations in the normal state provide informations about the
superconducting state itself. We shall show here that fluctuations enable to
detect the presence of a FFLO state in both 2D and 3D superconductors, and
allow to discriminate between the various a)-c) scenarios in the tetragonal 3D
case. We concentrate on the region near the tricritical point $(T^{\ast
},H^{\ast})$ which is the meeting point of the three transition lines
separating respectively the normal state, the uniform and the modulated
superconducting states \cite{saintjames,casalbuoni.nardulli.2004}.

In this paper, we evaluate the fluctuation induced magnetization near the FFLO
transition in both 2D and 3D anisotropic superconductors using the modified
Ginzburg-Landau (MGL) functional
\cite{buzdin_kachkachi(1997),yang_agterberg.2001,houzet2006}. Previously we
calculated the fluctuational specific heat and conductance near the pure FFLO
transition in the absence of orbital effect \cite{konschelle.press}. Our
motivation was to establish a relation between the topology of the lowest
energy fluctuation modes and the divergencies of the physical properties at
the FFLO transition. In the isotropic model, those divergencies are very
different than the standard BCS ones since the topologies of the degenerate
FFLO and BCS modes differ fundamentally. Unfortunalely, in the anisotropic
models, this degeneracy is lifted and the topologies of FFLO and BCS modes
become quite similar, thereby leading to less contrasted behaviors.

In the two-dimensional case, we also show that the ratio between the
paraconductivities along ($\sigma_{xx}$) and perpendicular ($\sigma_{yy}$) to
an in-plane applied magnetic field $\mathbf{H}=H\mathbf{e}_{x}$ is drastically
enhanced near the FFLO transition, in comparison to the one near the uniform
BCS transition. Moreover we demonstate that the high-field fluctuational
magnetization of thin films may oscillate between positive (paramagnetism) and
negative (diamagnetism) values. These oscillations originate from the
competition between orbital and paramagnetic effects which tend to promote
respectively Landau level modulation and FFLO modulation. Being precursors of
the Meisner or Abrikosov lattice state, the superconducting fluctuations are
usually diamagnetic. Therefore the paramagnetism predicted here is a hallmark
of the unconventional FFLO state. At lower field, the magnetization is
suppressed near the FFLO transition in comparison to the BCS case. These
features should also pertain in the case of layered 2D compounds like
$\lambda$-(BETS)$_{2}$FeCl$_{4}$ or $\kappa$-(BEDT-TTF)$_{2}$Cu(NCS)$_{2}$.

In 3D superconductors under high magnetic field, these oscillations are
blurred out when scenario a) is realized whereas they pertain when scenario b)
takes place, thereby providing an experimental test to distinghish among the
various possible structures of the order parameter described in
Ref.\cite{denisov09}. Experimentally, the superconducting fluctuations in
CeCoIn$_{5}$ have\ been investigated far above $T_{c\text{ }}$and under low
fields \cite{bel04,onose07}. Here we suggest similar measurements near
$T_{c\text{ }}$under strong magnetic field and near the FFLO critical temperature.

The paper is organized as follows. In Sec II, we present the MGL formalism
which includes higher order derivative of the order parameter than the
standard Ginzburg-Landau functional. Such an extension is necessary to handle
the nonuniform FFLO state. In Sec III, we analyse the case of thin
superconducting films under in-plane magnetic field and predict a strong
dependence of the conductance upon the mutual orientation of the current flow
and magnetic field. We also derive the fluctuation magnetization induced by a
tilted magnetic field pointing out of the film plane. In Sec IV, we discuss
the 3D anisotropic compounds with emphasis on the fluctuation magnetization.
In appendix, we provide a detailled derivation of the Ginzburg-Levanyuk
criterion for the FFLO transition in order to justify the Gaussian
approximation used in the whole paper.

\section{Formalism}

Here we introduce the\ modified Ginzburg-Landau (MGL) free-energy functional
and detail how the fluctuation induced properties can be obtained from the
spectrum of the fluctuations. This approach is valid near the tricritical
point. The location of the tricritical point in the temperature ($T$)-magnetic
field ($H$) phase diagram depends on microscopic details of the model like the
concentration and the type of impurities, the crystal and the order parameter
symmetries. Nevertheless the formula of this section are valid generically
around the tricritical point, independently of its precise location. Note that
for a clean $s$-wave superconductor the tricritical point is located at
$T^{\ast}=0.56T_{c0},\mu_{B}H^{\ast}=1.07k_{B}T_{c0}$, $T_{c0}$ being the
critical temperature in the absence of Zeeman effect and $\mu_{B}$ the Bohr
magneton \cite{saintjames,casalbuoni.nardulli.2004}.

\subsection{Modified Ginzburg-Landau functional}

At the vicinity of the tricritical point $(T^{\ast},H^{\ast})$, the FFLO
transition can be described by a Ginzburg-Landau like approach since the order
parameter $\Psi(\mathbf{r})$ and its spacial gradients are small
\cite{buzdin_kachkachi(1997)}. The corresponding free-energy functional,
called hereafter the modified Ginzburg-Landau (MGL) functional, differs from
the usual Ginzburg-Landau functional by the presence of higher order
derivatives of the order parameter $\Psi(\mathbf{r})$. This is related to the
fact that the FFLO phase corresponds to a non-uniform groundstate. This MGL
functional may be directly derived from the microscopic BCS theory for clean
isotropic $s$-wave superconductors with Zeeman interaction
\cite{buzdin_kachkachi(1997)}, and allows extention to conventional and
unconventional singlet superconductors in the presence of paramagnetic,
orbital and impurity effects \cite{yang_agterberg.2001,houzet2006}. The
quadratic terms ($\left\vert \Psi(\mathbf{r})\right\vert ^{2}$, $\left\vert
\mathbf{\nabla}\Psi(\mathbf{r})\right\vert ^{2}$, etc...) of the MGL describe
the free dynamics of the order parameter while higher-order terms ($\left\vert
\Psi(\mathbf{r})\right\vert ^{4}$, $\left\vert \Psi(\mathbf{r})\right\vert
^{6}$ etc...) account for the interactions. In this paper, we analyse the
fluctuations within the Gaussian approximation whereby only the quadratic
terms are retained \cite{b.landau.V_e}. In the appendix, we shall justify this
approximation by demonstrating that the interaction terms are small in the
experimentally relevant range of parameter $T_{c0}/E_{F}\ll1$, $E_{F}$ being
the Fermi energy.

Within the Gaussian approximation, the MGL functional writes
\begin{equation}
H_{G}\left[  \Psi\right]  =\int d\boldsymbol{r}\left(  \alpha\left\vert
\Psi\right\vert ^{2}-g_{i}\left\vert D_{i}\Psi\right\vert ^{2}+\gamma
_{ij}\left\vert D_{i}D_{j}\Psi\right\vert ^{2}\right)  , \label{equation1}%
\end{equation}
where the summation over the index $i=x,y,z$ is implied, and $\alpha=a\left(
T-\tilde{T}_{c}\right)  $, $\tilde{T}_{c}\left(  H\right)  $ being the
critical temperature for the uniform superconducting second order phase
transition. The components $A_{i}$ of the vector potential enter the MGL
through the covariant derivatives $D_{i}=\partial_{i}+2ieA_{i}/\hslash$, while
the coefficients $\alpha,g_{i}\mathbf{\ }$and $\gamma_{ij}$ are functions of
the temperature and field. The detailled microscopic expressions of these
coefficients as functions of $T$ and $H$ can be found in
\cite{buzdin_kachkachi(1997),yang_agterberg.2001,houzet2006} for various
crystal lattices and in presence of paramagnetic, orbital and impurity
effects. As a salient and common feature of these functionals
\cite{buzdin_kachkachi(1997),yang_agterberg.2001,houzet2006}, coefficients
$g_{i}$ change sign at the tricritical point, thereby inducing an
inhomogeneous superconducting phase when $g_{i}$ $>0$, namely at low
temperature and high field ($H/T>H^{\ast}/T^{\ast}$). Then the FFLO critical
temperature $T_{c}\left(  H\right)  $ is larger than $\tilde{T}_{c}\left(
H\right)  $ implying a transition between the normal state and a nonuniform
superconducting state. Note that the idealized isotropic form of this
functional corresponds to $g_{i}=g$ and $\gamma_{ij}=\gamma$.

\subsection{Fluctuation free energy in the absence of orbital effect}

In the absence of magnetic field, the order parameter can be expanded in plane
waves as $\Psi(\mathbf{r})=\sum_{\mathbf{k}}\Psi_{\mathbf{k}}e^{i\mathbf{k.r}%
}$. In this Fourier representation, the free-energy Eq. ($\ref{equation1}$)
can be rewritten as $H_{G}\left[  \Psi\right]  =\sum_{\mathbf{k}}%
\varepsilon_{\mathbf{k}}\left\vert \Psi_{\mathbf{k}}\right\vert ^{2}$ where
\begin{equation}
\varepsilon_{\mathbf{k}}=\alpha-g_{i}k_{i}^{2}+\gamma_{ij}k_{i}^{2}k_{j}^{2}
\label{spectrumpara}%
\end{equation}
describes the spectrum of the decoupled fluctuation modes $\Psi_{\mathbf{k}}$.
The partition function $Z=Tr(e^{-H_{G}/k_{B}T})$ is obtained by tracing over
all the possible values of these modes $\Psi_{\mathbf{k}}$ which reduces to an
infinite product of Gaussian integrals:%
\begin{equation}
Z=%
{\textstyle\prod\limits_{\mathbf{k}}}
\frac{\pi k_{B}T}{\varepsilon_{\mathbf{k}}}.
\end{equation}
The corresponding thermodynamical free energy (per unit volume) $F=-k_{B}T\ln
Z$ is given by\cite{skocpol}%
\begin{equation}
F=k_{B}T\int\frac{d\mathbf{k}^{d}}{(2\pi)^{d}}\ln\frac{\varepsilon
_{\mathbf{k}}}{\pi k_{B}T}.
\end{equation}
In the isotropic case ($g_{i}=g$ and $\gamma_{ij}=\gamma$), the spectrum
$\varepsilon_{\mathbf{k}}=\alpha-g\mathbf{k}^{2}+\gamma\mathbf{k}^{4}$ can be
exactly rewritten as
\begin{equation}
\varepsilon_{\mathbf{k}}=\tau+\gamma\left(  \mathbf{k}^{2}-q_{0}^{2}\right)
^{2} \label{spectrumiso}%
\end{equation}
where $q_{0}^{2}=g/2\gamma$ and $\tau=\alpha-g^{2}/4\gamma$. The FFLO
transition arises at the critical temperature $T_{c}\left(  H\right)  $
defined by $\tau=0$, namely at:
\begin{equation}
T_{c}\left(  H\right)  =\tilde{T}_{c}\left(  H\right)  +\frac{g^{2}}{4\gamma
a}, \label{TCiso}%
\end{equation}
which is higher than the critical temperature for the transition towards a
uniform superconductor $\tilde{T}_{c}\left(  H\right)  $. In the normal state
$T>T_{c}\left(  H\right)  $, Eq. ($\ref{spectrumiso}$) makes apparent that the
lowest energy fluctuation modes are degenerate and located around the sphere
$\mathbf{k}^{2}=q_{0}^{2}$ in reciprocal space.

\subsection{Fluctuation free energy with orbital effect}

Here we consider Eq. ($\ref{equation1}$) in the isotropic case ($g_{i}=g$ and
$\gamma_{ij}=\gamma$) in presence of the orbital effect associated with a
magnetic field $\mathbf{H}=H\mathbf{e}_{z}$. Within the gauge $A_{x}=0$ and
$A_{y}=xH$, the order parameter is conveniently represented in terms of Landau
wavefunctions. This follows from the fact that the mean-field equation $\delta
H/\delta\Psi^{\ast}=0$ reads
\begin{equation}
(\alpha+g\mathbf{D}^{2}+\gamma\mathbf{D}^{4})\Psi(x,y,z)=E\Psi(x,y,z).
\label{meanfield}%
\end{equation}
Owing to the translational invariance along $y$ and $z$, the momenta $k_{y}$
and $k_{z}$ are good quantum numbers. In the absence of the fourth-order
derivative ($\gamma=0$), the solutions of Eq. ($\ref{meanfield}$) are well
known \cite{Landau3}, being the standard Landau wavefunctions $\Psi
(\mathbf{r})=\Phi_{n,k_{y},k_{z}}=f(x)e^{ik_{y}y+ik_{z}z}$. The equation for
$f(x)$
\begin{equation}
g\frac{d^{2}f}{dx^{2}}-g\left(  k_{y}+\frac{2e}{\hbar}Hx\right)  ^{2}%
f+(\alpha-gk_{z}^{2})f=Ef(x)
\end{equation}
is similar to the harmonic oscillator equation with "inverse mass" $-g$ and
the frequency $\omega_{c}=-4eHg/\hbar^{2}$. Therefore the Landau levels for
($\gamma=0$) are given by \cite{Landau3}
\begin{equation}
E_{n}(k_{z})=\alpha-gk_{z}^{2}+\hslash\omega_{c}(n+\frac{1}{2})
\end{equation}
where $n=0,1,2,..$ while $k_{z}$ is a continuous wavevector.

We now solve Eq. ($\ref{meanfield}$) in the presence of the fourth-order
derivative ($\gamma\neq0$) which is a hallmark of the MGL functional and FFLO
state. We first show that the Landau wavefunctions $\Phi_{n,k_{y},k_{z}%
}(\mathbf{r})$ (eigenstates of $\alpha+g\mathbf{D}^{2}$) are still eigenstates
of the differential operator $\alpha+g\mathbf{D}^{2}+\gamma\mathbf{D}^{4}$.
Indeed introducing the quantized wavevector $Q_{n}$ (\ $n=0,1,2,...$) defined
by
\begin{equation}
Q_{n}^{2}\equiv\frac{\hslash\omega_{c}}{-g}(n+\frac{1}{2})=\dfrac{4eH}%
{\hslash}\left(  n+\dfrac{1}{2}\right)  \text{,}%
\end{equation}
one obtains immediately that
\begin{align}
g\mathbf{D}^{2}\Phi_{n,k_{y},k_{z}}  &  =-g\left(  k_{z}^{2}+Q_{n}^{2}\right)
\Phi_{n,k_{y},k_{z}},\\
\gamma\mathbf{D}^{4}\Phi_{n,k_{y},k_{z}}  &  =\gamma(k_{z}^{2}+Q_{n}^{2}%
)^{2}\Phi_{n,k_{y},k_{z}},
\end{align}
and therefore\cite{footnoteLandau}
\begin{equation}
(\alpha+g\mathbf{D}^{2}+\gamma\mathbf{D}^{4})\Phi_{n,k_{y},k_{z}}=E_{n}%
(k_{z})\Phi_{n,k_{y},k_{z}},
\end{equation}
with the fluctuation spectrum%
\begin{equation}
E_{n}(k_{z})=\tau+\gamma(Q_{n}^{2}+k_{z}^{2}-q_{0}^{2})^{2}.
\label{spectreLandau}%
\end{equation}
Note that $E_{n}(k_{z})$ is still degenerate, being independent of $k_{y}$.
Finally the free energy functional reduces to the sum
\begin{equation}
H_{G}\left[  \Psi\right]  =\sum_{n,k_{y},k_{z}}E_{n}(k_{z})\left\vert
\Phi_{n,k_{y},k_{z}}\right\vert ^{2}%
\end{equation}
over decoupled modes $\Phi_{n,k_{y},k_{z}}$. The partition function
$Z=Tr(e^{-H_{G}/k_{B}T})$ is obtained by tracing over all the possible values
of these modes $\Phi_{n,k_{y},k_{z}}$ and finally reduces to an infinite
product of Gaussian integrals:%
\begin{equation}
Z=%
{\textstyle\prod\limits_{n,k_{y},k_{z}}}
\frac{\pi k_{B}T}{E_{n}(k_{z})}.
\end{equation}
The corresponding thermodynamical free energy (per unit volume) $F=-k_{B}T\ln
Z$ is given by%
\begin{equation}
F=\frac{H}{\Phi_{0}}k_{B}T\overset{\infty}{\sum_{n=0}}\int\frac{dk_{z}}{2\pi
}\ln\frac{E_{n}(k_{z})}{\pi k_{B}T}, \label{F3D}%
\end{equation}
where the prefactor $H/\Phi_{0}$ accounts for the degeneracy of each Landau
level at given $n$ and $k_{z}$ \cite{Landau3}. Here $\Phi_{0}=h/2e$ is the
superconducting flux quantum. In the two-dimensional case the quantum number
$k_{z}$ is irrelevant and the average free energy per unit surface is given by%

\begin{equation}
F=\frac{H}{\Phi_{0}}k_{B}T\overset{\infty}{\sum_{n=0}}\ln\frac{E_{n}(k_{z}%
=0)}{\pi k_{B}T}. \label{F2D}%
\end{equation}
The magnetization along the $z$-axis is simply given by $M=-\partial
F/\partial H$.

\subsection{Transport}

Besides thermodynamics, the fluctuations also affect the transport properties.
A standard procedure consists in using the time-dependent Ginzburg-Landau
equation to obtain the current-current correlator at different times
\cite{tinkham,skocpol}. Then a general formula for the paraconductivity
tensor
\begin{equation}
\sigma_{ij}=\frac{\pi e^{2}ak_{B}T}{4\hslash}\int\frac{d\mathbf{k}^{d}}%
{(2\pi)^{d}}\frac{v_{\mathbf{k}i}v_{\mathbf{k}j}}{\varepsilon_{\mathbf{k}}%
^{3}} \label{paraconductivity}%
\end{equation}
can be obtained within the Kubo formalism \cite{b.larkin_varlamov}. It should
be noticed that the momentum dependence of the velocity component
$v_{\mathbf{k}i}=\partial\varepsilon_{\mathbf{k}}/\partial k_{i}$ along the
$i-$axis is very different from its usual form owing to the presence of the
$\gamma_{ij}k_{i}^{2}k_{j}^{2}$ terms in the dispersion relation
Eq.($\ref{spectrumpara}$). In particular, $v_{\mathbf{k}i}$ are no longer
linear combinaisons of the momentum components $k_{i}$. The formula
Eq.($\ref{paraconductivity}$) provides the so-called classical
Aslamasov-Larkin contribution to the paraconductivity. It is well known that
the quantum Maki-Thomson contribution can be important especially in the
two-dimensional case \cite{b.larkin_varlamov}. In this paper, we study the
FFLO transition at the vicinity of the tricritical point $(T^{\ast},H^{\ast})$
where the strong pair-breaking mechanism suppresses the Maki-Thomson contribution.

\section{Superconducting thin films}

Here we consider thin superconducting films where the FFLO state is realized
due to the paramagnetic effect of an in-plane magnetic field\ $\mathbf{H}%
=H_{\parallel}\mathbf{e}_{x}$. We assume that the lattice has the square
symmetry and thus equivalent properties in the $x$ and $y$ directions in the
absence of field. In Sec III.A, we first neglect the orbital effect associated
with $\mathbf{H}=H_{\parallel}\mathbf{e}_{x}$ and calculate the
paraconductivity $\sigma_{xx}=\sigma_{yy}$. Then we treat perturbatively the
orbital effect associated with $\mathbf{H}=H_{\parallel}\mathbf{e}_{x}$, and
find that the paraconductivity $\sigma_{xx}$ measured along the applied
magnetic field differs from the one ($\sigma_{yy}$) measured perpendicular to
the field (Sec. III.B). Finally, we also discuss the effect of a perpendicular
magnetic field $\mathbf{H}=H\mathbf{e}_{z}$ which induces magnetization
oscillations between diamagnetism and paramagnetism (Sec. III.C). This
behavior is in sharp contrast with the usual fluctuation induced diamagnetism
predicted and observed close to the BCS
transition\cite{skocpol,b.larkin_varlamov}.

\subsection{Pure paramagnetic
limit\label{PAR_explicit_derivation_cubic_anisotropy}}

\begin{figure}[ptb]
\begin{center}
\includegraphics[scale=0.4,angle=0]{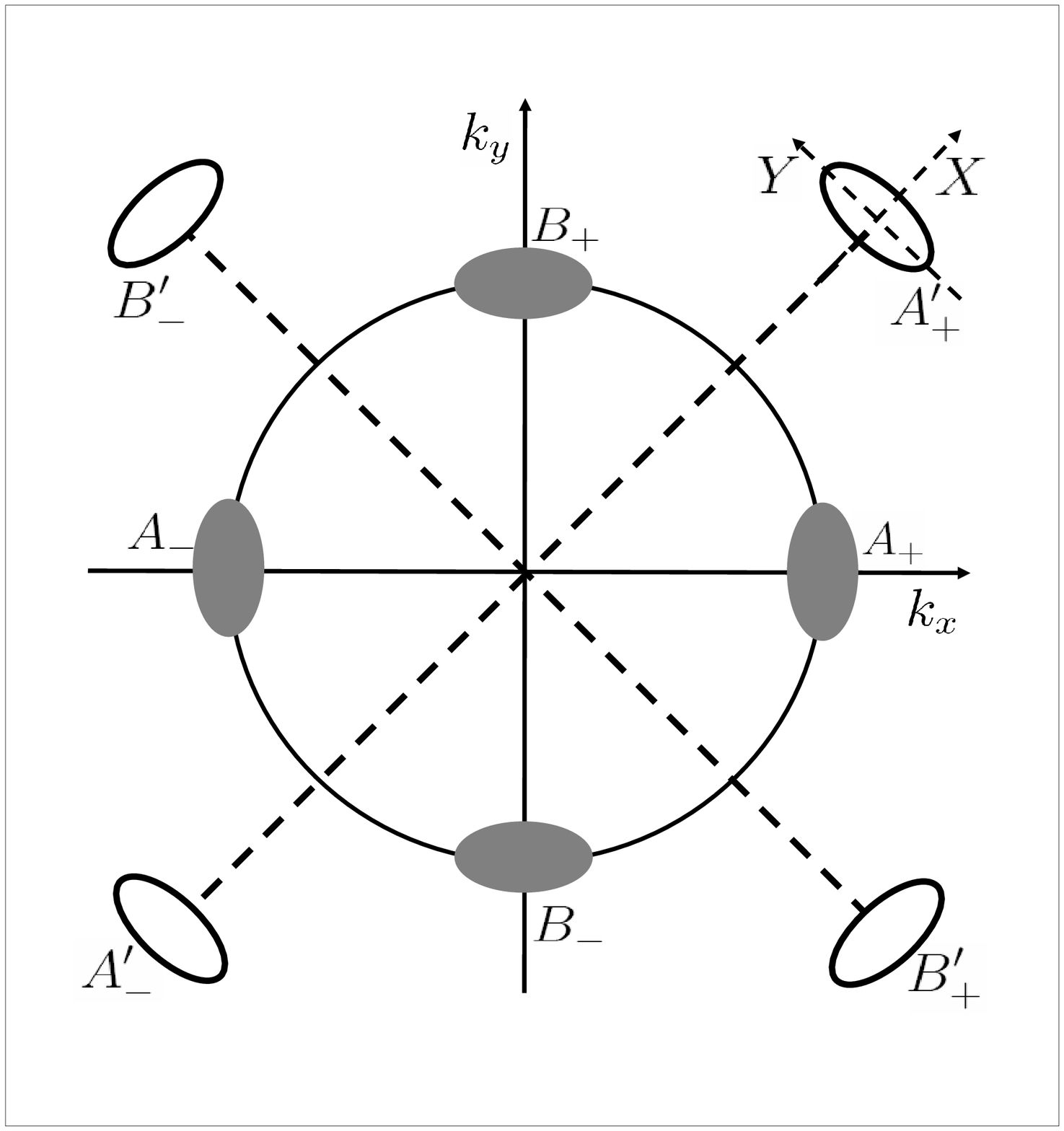}
\end{center}
\caption{Minima of the fluctuation spectrum $\varepsilon_{\mathbf{k}}$ for the
two-dimensional model with square symmetry. The filled (open) ellipses
indicate the locations of the lowest energy fluctuations for $\eta>1$
($\eta<1$). Note that in the isotropic model ($\eta=1$), there is an infinity
of degenerate minima located on the circle (solid line). }%
\label{FIG_cubic_anisotropy}%
\end{figure}

In the pure paramagnetic limit, the in-plane magnetic field $\mathbf{H}%
=H_{\parallel}\mathbf{e}_{x}$ only acts on the spins.\ Hence one can set
$A_{i}=0$ in the MGL functional Eq.($\ref{equation1}$) and use the spectrum
Eq.($\ref{spectrumpara}$) to describe the superconducting fluctuations in the
normal state. By increasing $H_{\parallel}$ and lowering the temperature, one
may tune the thin film near the tricritical point where $g=0$. The parameter
$g$ may change sign, thereby indicating an instability towards the
inhomogeneous FFLO state when $g$ is positive.

Furthermore the square symmetry implies $g_{x}=g_{y}=g$, $\gamma_{xx}%
=\gamma_{yy}=\gamma$ and $\gamma_{xy}=\gamma_{yx}=\eta\gamma$ in
Eq.($\ref{spectrumpara}$). Isotropy is only restored when $\eta=1$ which
corresponds to a circular Fermi surface \cite{konschelle.press}. Then the low
energy modes are located around the whole circle defined by $\mathbf{k}%
^{2}=q_{0}^{2}=g/2\gamma$ in reciprocal space, according to
Eq.($\ref{spectrumiso}$). In the general case\ ($\eta\neq1$), the quartic
terms $\gamma_{ij}k_{i}^{2}k_{j}^{2}$ introduce a non trivial anisotropy, and
the lowest energy are realized around isolated points of the reciprocal state
(Fig.\ref{FIG_cubic_anisotropy}).

The paraconductivity $\sigma_{xx}=\sigma_{yy}$ near the FFLO transition
($g>0$) is obtained from Eq.($\ref{paraconductivity}$) where the integral
extends over the whole reciprocal space. Nevertheless due to the
denominator\ $\varepsilon_{\mathbf{k}}^{3}$ in Eq.($\ref{paraconductivity}$),
the main contribution comes from the lowest energy fluctuation modes.

We shall first identify these isolated minima $\mathbf{k}_{\min}$ of the
energy by solving $\partial\varepsilon_{\mathbf{k}}/\partial k_{x}%
(\mathbf{k}_{\min})=\partial\varepsilon_{\mathbf{k}}/\partial k_{y}%
(\mathbf{k}_{\min})=0$. The location of these minima differs in the cases
$\eta>1$ and $\eta<1$ respectively (Fig.\ref{FIG_cubic_anisotropy}). In the
case $\eta>1$, the spectrum $\varepsilon_{\mathbf{k}}$ has four minima located
along the $x$ and $y$ axis, namely at points $A_{\pm}\left(  \pm
q_{0},0\right)  $ and $B_{\pm}\left(  0,\pm q_{0}\right)  $ of the reciprocal
space (Fig.\ref{FIG_cubic_anisotropy}). Near the minima $A_{\pm}$ located
along the $x$ axis, we find that the spectrum can be approximated by
\begin{equation}
\varepsilon_{\mathbf{k}}^{(A_{\pm})}\approx\tau+2g\left(  k_{x}\mp
q_{0}\right)  ^{2}+g\left(  \eta-1\right)  k_{y}^{2} \label{specA}%
\end{equation}
with $\tau=\alpha-g^{2}/4\gamma=a\left(  T-T_{c}\right)  $. The expression for
the FFLO critical temperature $T_{c}$ coincides with Eq.($\ref{TCiso}$) as
long as $\eta>1$.

Similarly, the spectrum around the minima $B_{\pm}\left(  0,\pm q_{0}\right)
$ located along the $y$ axis is given by
\begin{equation}
\varepsilon_{\mathbf{k}}^{(B_{\pm})}\approx\tau+g\left(  \eta-1\right)
k_{x}^{2}+2g\left(  k_{y}\mp q_{0}\right)  ^{2}. \label{specB}%
\end{equation}

The next step consists in evaluating the generalized velocities around each
minima. For instance around $A_{+}$, we obtain
\begin{align}
v_{\mathbf{k}x}  &  =\frac{\partial\varepsilon_{\mathbf{k}}^{(A_{+})}%
}{\partial k_{x}}=4g\left(  k_{x}-q_{0}\right)  ,\\
v_{\mathbf{k}y}  &  =\frac{\partial\varepsilon_{\mathbf{k}}^{(A_{+})}%
}{\partial k_{y}}=2g\left(  \eta-1\right)  k_{y}.
\end{align}
Finally, we evaluate the integral Eq.($\ref{paraconductivity}$) around the
minimum $A_{+}$.

Reproducing this steps for the other minima $A_{-},B_{+}$ and $B_{-}$ and
summing the contributions from the four minima, one obtains the
paraconductivity
\begin{equation}
\sigma_{xx}=\sigma_{yy}=\frac{e^{2}ak_{B}}{4\sqrt{2}\hslash}\frac{1+\eta
}{\sqrt{\eta-1}}\left(  \frac{T_{c}}{T-T_{c}}\right)  \text{.}
\label{EQ_square_example_out_of_diagonals}%
\end{equation}
which diverges at the FFLO transition.

The second case $\eta<1$ can be treated along the same line of reasoning,
albeit the four degenerate minima $\mathbf{k}_{\min}^{\prime}=(k_{\min
x}^{\prime},k_{\min y}^{\prime})$ are now located on the diagonals with
$(k_{\min x}^{\prime})^{2}=(k_{\min y}^{\prime})^{2}=q_{0}^{2}/2$
(Fig.\ref{FIG_cubic_anisotropy}). The spectrum can be expanded as
\begin{equation}
\varepsilon_{\mathbf{k}}\approx\tau-%
{\textstyle\sum\limits_{i,j=x,y}}
\lambda_{ij}\left(  k_{i}-k_{\min i}^{\prime}\right)  \left(  k_{j}-k_{\min
j}^{\prime}\right)  ,
\end{equation}
around any of those four minima denoted $A_{+}^{\prime},A_{-}^{\prime}%
,B_{+}^{\prime}$ and $B_{-}^{\prime}$ (Fig.\ref{FIG_cubic_anisotropy}), with
$\tau=\alpha-g^{2}/2\gamma\left(  1+\eta\right)  =a\left(  T-T_{c}\right)  $.
For instance, we find
\begin{equation}
\lambda_{xx}=\lambda_{yy}=\frac{2g}{1+\eta}\text{ \ and \ }\lambda
_{xy}=\lambda_{yx}=\frac{2g\eta}{1+\eta}%
\end{equation}
around $A_{+}^{\prime}(q_{0}/\sqrt{2},q_{0}/\sqrt{2})$. Diagonalization of the
tensor $\lambda_{ij}$ leads to the eigenvalues $\lambda_{X}=2g$ and
$\lambda_{Y}=2g\left(  1-\eta\right)  /\left(  1+\eta\right)  $ along the
principal axis $X$ and $Y$.

Finally summation over the four minima (Fig.\ref{FIG_cubic_anisotropy})
yields:
\begin{equation}
\sigma_{xx}=\sigma_{yy}=\frac{e^{2}ak_{B}}{2\hslash}\frac{g^{2}\eta^{2}%
}{\left(  1+\eta\right)  ^{3}}\sqrt{\frac{1+\eta}{1-\eta}}\left(  \frac{T_{c}%
}{T-T_{c}}\right)  . \label{EQ_square_example_on_diagonals}%
\end{equation}

Note that in this regime $\eta<1$, the expression for the FFLO critical
temperature $T_{c}$ differs slightly from Eq.($\ref{TCiso}$).

The above expressions Eqs.$\left(  \ref{EQ_square_example_out_of_diagonals}%
\text{,}\ref{EQ_square_example_on_diagonals}\right)  $ both diverge for
$\eta\rightarrow1$, which indicates stronger Gaussian fluctuations in the
isotropic model \cite{konschelle.press}.

\subsection{Orbital effect associated with an in-plane magnetic field
\label{PAR_anisotropic_H}}

We now take into account the orbital effect associated with the in-plane
magnetic field ($\mathbf{H}=H_{\parallel}\mathbf{e}_{x}$), and show that it
breaks the square symmetry, inducing distinct paraconductivities along $x$ and
$y$ directions.

In the case of thin films with a strong confinement in the $z$ direction, this
orbital effect is small. Then it is still possible to describe the
fluctuations by the spectrum Eq.($\ref{spectrumpara}$) with $H_{\parallel}%
$-dependent coefficients \cite{buzdin_matsuda_shibauchi.2007}, namely by
\begin{multline}
\varepsilon_{\mathbf{k}}=\alpha-g\left(  k^{2}+\dfrac{\left(  H_{\parallel
}d\right)  ^{2}}{12\Phi_{0}^{2}}\right)  +\gamma\left(  k^{4}+\dfrac{\left(
H_{\parallel}d\right)  ^{4}}{80\Phi_{0}^{4}}\right)  +\\
+2\gamma\left(  \eta-1\right)  k_{x}^{2}k_{y}^{2}+\dfrac{\left(  H_{\parallel
}d\right)  ^{2}}{6\Phi_{0}^{2}}\left(  \gamma\eta k_{x}^{2}+3\gamma k_{y}%
^{2}\right)  , \label{EQ_functionnal_with_H}%
\end{multline}
where $k^{2}=k_{x}^{2}+k_{y}^{2}$, $d$ is the width of the film along the
$z$-axis, and $\Phi_{0}=h/2e$ is the superconducting quantum of flux. Owing to
the smallness of the dimensionless parameter $(H_{\parallel}\xi d/q_{0}%
\Phi_{0})^{2}$, it is still possible to use Eq.($\ref{paraconductivity}$) in
order to evaluate the paraconductivity tensor following the same procedure
than in the previous section \ref{PAR_explicit_derivation_cubic_anisotropy}.
Here $\xi$ is the superconducting coherence length.

Due to the field ($H_{\parallel}$) dependence of the coefficients in
Eq.($\ref{EQ_functionnal_with_H}$), the minima of $\varepsilon_{\mathbf{k}}$
are displaced (in comparison to the case $H_{\parallel}\xi d/q_{0}\Phi_{0}=0$
shown in Fig. 1) according to ($\eta>1$):%
\begin{align}
A_{\pm}  &  \rightarrow A_{\pm}\left(  \pm\sqrt{\frac{g}{2\gamma}-\frac
{\eta\left(  H_{\parallel}d\right)  ^{2}}{12\Phi_{0}^{2}}},0\right)  ,\\
B_{\pm}  &  \rightarrow B_{\pm}\left(  0,\pm\sqrt{\frac{g}{2\gamma}%
-\frac{\left(  H_{\parallel}d\right)  ^{2}}{4\Phi_{0}^{2}}}\right)  .
\end{align}
Moreover the square symmetry is broken since the critical temperature
associated with modulation $A_{\pm}$ differs from the one for $B_{\pm}$. It
happens that the FFLO modulation occurs along the field (points $A_{\pm}$) for
$\eta<3$. In contrast for $\eta>3$ the modulation occurs along the $y$-axis
(points $B_{\pm}$) which is perpendicular to the applied field
\cite{buzdin_matsuda_shibauchi.2007}.

We concentrate on the case for $1<\eta<3$ wherein the order parameter is
modulated along the field $\mathbf{H}=H_{\parallel}\mathbf{e}_{x}$. The
paraconductivity comes from the contributions around the points $A_{+}$ and
$A_{-}$. Then the paraconductivity $\sigma_{xx}$ measured along the field
differs from the one $\sigma_{yy}$ measured in the perpendicular direction. As
a main result of this section, the ratio $\sigma_{xx}/\sigma_{yy}$ contains a
contribution which diverges at the FFLO transition ($g>0$)
\begin{equation}
\left(  \frac{\sigma_{xx}}{\sigma_{yy}}\right)  =1+\frac{\gamma}{6g}\left(
\frac{H_{\parallel}d}{\Phi_{0}}\right)  ^{2}\left[  \frac{\eta-3}{\eta
-1}+\frac{g^{2}}{2\gamma\tau}\frac{(\eta-3)^{2}}{\eta+1}\right]  .
\end{equation}
This term produces a strong enhancement of \ $\sigma_{xx}/\sigma_{yy}$ when
the transition is approached, namely when $\tau=a\left(  T-T_{c}\right)
\rightarrow0^{+}$. This is in sharp contrast with the transition towards a
uniform BCS state. There the ratio of the paraconductivities
\begin{equation}
\left(  \frac{\sigma_{xx}}{\sigma_{yy}}\right)  _{BCS}=1-\frac{\gamma\left(
\eta-3\right)  }{6g}\left(  \frac{H_{\parallel}d}{\Phi_{0}}\right)  ^{2}%
\end{equation}
does not contain any divergent term at the BCS transition ($g<0$). Such an
enhancement of $\sigma_{xx}/\sigma_{yy}$ may serve as an experimental
signature of the FFLO state. This property is reminiscent of the recent
finding that critical current oscillates as a function of magnetic orientation
in anisotropic 2D films \cite{buzdin_matsuda_shibauchi.2007}.

\subsection{Orbital effect in a perpendicular magnetic field
\label{PAR_fluct_magnetic_susceptibility}}

Finally, we discuss the effect of a perpendicular component $H_{\perp
}\boldsymbol{e}_{z}$ when the field $\mathbf{H}=H_{\parallel}\boldsymbol{e}%
_{x}+H_{\perp}\boldsymbol{e}_{z}$ is tilted out of the film plane. The
perpendicular component $H_{\perp}\boldsymbol{e}_{z}$ quantizes the in-plane
motion of the fluctuating Cooper pairs, and induces a finite magnetization.
This effect is larger than \ the orbital motion associated with the in-plane
part of the field. We therefore neglect the later (which is correct for
$H_{\parallel}\xi d\ll\Phi_{0}$) and use the MGL functional
Eq.($\ref{equation1}$) within the gauge $A_{x}=0$ and $A_{y}=xH_{\perp}$, like
in section II.C.

In the anisotropic case, the eigenmodes of this functional are not known
exactly for finite $H_{\perp}$ precluding an analytical evaluation of the
magnetization. Nevertheless the isotropic model already exhibits oscillations
of the fluctuation magnetization. Recently the fluctuational magnetization
(persistent current) of small rings made of a FFLO superconductor was obtained
within the framework of the isotropic model introduced in Sec. II.C. In the
following, we derive a simple formula for the magnetization in the simpler
planar geometry of superconducting films \cite{zyuzin09}.

The fluctuation spectrum is given by Eq.($\ref{spectreLandau}$) and the free
energy per unit surface by Eq. ($\ref{F2D}$) with $H=H_{\perp}$. The
particular form Eq.($\ref{spectreLandau}$) of the spectrum enables both
degeneracies between the Landau levels ($E_{n}=E_{n+1}$) and commensurability
effects between the wavevectors $Q_{n}$ and $q_{0}$.

\textit{Single mode (high fields) approximation:} For large perpendicular
field, namely $H_{\perp}/\Phi_{0}\gg$ $\sqrt{\tau/\gamma}$, the Landau levels
are well separated from each others and the main contribution to the free
energy Eq.$\left(  \ref{F2D}\right)  $ comes either from the single level with
minimal energy $E_{n}$, or from two levels when a degeneracy ($E_{n}=E_{n+1}$) occurs.

Let us first consider the non-degenerate case. Then the free energy is simply
given by the single level contribution
\begin{equation}
F_{n}=\frac{H_{\perp}}{\Phi_{0}}k_{B}T\ln\frac{E_{n}(k_{z}=0)}{\pi k_{B}T}%
\end{equation}
and the corresponding orbital magnetization (per unit surface)
\begin{equation}
M_{n}=-\frac{(8n+4)k_{B}T}{\Phi_{0}^{2}}\left(  \frac{\gamma(Q_{n}^{2}%
-q_{0}^{2})}{\tau+\gamma(Q_{n}^{2}-q_{0}^{2})^{2}}\right)  H_{\perp}
\label{Eq_Msingle}%
\end{equation}
is highly nonlinear since the prefactor of $H_{\perp}$ depends strongly on the
field and on temperature. Importantly the magnetization may change sign due to
the presence of the factor $Q_{n}^{2}-q_{0}^{2}$ in the numerator. In order to
make more transparent the formula Eq.$\left(  \ref{Eq_Msingle}\right)  $, one
may introduce the field-dependent temperature $T_{cn}(H)$ where the
denominator vanishes:%
\begin{equation}
a(T-T_{cn})=\tau+\gamma(Q_{n}^{2}-q_{0}^{2})^{2}%
\end{equation}
This relation defines the second-order transition line $T_{c}^{(n)}(H)$
between the normal and the modulated superconducting state described by the
$n$-th Landau level. We also define the points like A,C,E
(Fig.\ref{FIG_phase_diagram_anisotropic_functional}) along this transition
line where the numerator vanishes since $Q_{n}^{2}=q_{0}^{2}$. Those points
are also located on the second-order transition line $T_{cP}(H)$ between the
normal and the FFLO superconducting state calculated in the pure paramagnetic
limit. In the normal state, the orbital magnetization can be therefore
reexpressed as%
\begin{equation}
M_{n}=-\frac{(8n+4)k_{B}T}{\Phi_{0}^{2}}\frac{\gamma}{a}\left(  \frac
{Q_{n}^{2}-q_{0}^{2}}{T-T_{cn}}\right)  H_{\perp}%
\end{equation}
This 2D magnetization is diamagnetic when $Q_{n}^{2}>q_{0}^{2}$ and
paramagnetic when $Q_{n}^{2}<q_{0}^{2}$
(Fig.\ref{FIG_phase_diagram_anisotropic_functional}). In contrast, the
fluctuation magnetization is always diamagnetic in the BCS case. However
$M_{n}$ follows a similar power law $(T-T_{c}^{(n)})^{-1}$ and is on the same
order of magnitude than the BCS magnetization \cite{b.larkin_varlamov}.
Consequently we expect that the oscillations between diamagnetism and
paramagnetism should be measurable in thin films of FFLO superconductors. This
single mode approximation breaks down when the $n$-th and ($n+1$)-th Landau
levels are degenerate, namely when $E_{n}=E_{n+1}$. Then the two levels must
be included together in the free energy, whereas the other Landau levels are
still far in energy and can be neglected safely. The resulting magnetization
$M_{n}+M_{n+1}$ is slightly diamagnetic at degeneracy.

\textit{Continuum (low fields) approximation: }The single mode approximation
breaks down in the weak field limit ($H_{\perp}/\Phi_{0}\ll$ $\sqrt
{\tau/\gamma}$) where the Landau level separation becomes so small that all
the levels have to be taken into account. This situation corresponds to a
magnetic field which is slightly tilted out of the film plane. In the case of
a uniform BCS superconductor, the standard result for the magnetization is
\cite{schmid.1969}:
\begin{equation}
M=\frac{\pi}{3}\frac{k_{B}T}{\Phi_{0}^{2}}\frac{g}{a(T-\tilde{T}_{c})}%
H_{\perp} \label{divergency}%
\end{equation}
which is diamagnetic ($g<0$) and diverges at the BCS\ critical temperature
$\tilde{T}_{c}\left(  H\right)  $ for the second order phase transition. This
diamagnetic response is suppressed when the tricritical point is approached,
i.e. when $g\rightarrow0^{-}$
(Fig.\ref{FIG_phase_diagram_anisotropic_functional})\cite{konschelle.press}.
On the FFLO side (region $g>0$ in
Fig.\ref{FIG_phase_diagram_anisotropic_functional}) and for a given magnetic
field, the system becomes a FFLO superconductor before the divergency develops
because $T_{c}>\tilde{T}_{c}$.

\textit{Conclusion:} At high fields $H_{\perp}$, the magnetization near the
FFLO transition line oscillates between sizeable diamagnetism and
paramagnetism as the transitions between successive Landau levels are
realized. In the low field limit, these transitions become very close and the
oscillations average themselves leading to a cancellation of the linear
response and a suppression of the divergency. This situation is in strong
contrast with the standard BCS case where the magnetization is always
diamagnetic. We suggest to perform magnetization measurements in thin films
near an expected FFLO transition. The suppression of the fluctuational
magnetization at low perpendicular field and the restoration of sizeable
oscillations between para- and diamagnetism at higher perpendicular field
should be a strong indication for the FFLO state in quasi-two dimensional
compounds.\bigskip

\section{Anisotropic 3D superconductors\label{PAR_3D_systems}}

It is commonly believed that the FFLO state in CeCoIn$_{5}$ corresponds to a
modulation along the applied magnetic field. Nevertheless it was argued
recently that this situation is unlikely to happen for arbitrary field
orientations when the tetragonal anisotropy of CeCoIn$_{5}$ is properly taken
into account. Apparently if the order parameter modulation is along the field
for $\mathbf{H\perp c}$ (resp. $\mathbf{H\parallel c}$), then the modulation
is likely to be perpendicular to the field for $\mathbf{H\parallel c}$ (resp.
$\mathbf{H\perp c}$) \cite{denisov09}. Here we investigate the FFLO
fluctuations in anisotropic 3D compounds, building upon the various mean-field
scenarios reported in Ref\cite{denisov09}. We evaluate the fluctuational
magnetization $M$ along the magnetic field $\mathbf{H}=H\mathbf{e}_{z}$. In
particular, we demonstrate below that the magnetization oscillates between
sizeable diamagnetism and paramagnetism when the modulation is perpendicular
to the field (Landau level like). Those oscillations are the 3D counterparts
of the ones predicted in the previous section for superconducting films. In
contrast the magnetization is shown to be strongly suppressed when the
modulation occurs along the field (FFLO like modulation). In the 3D case,
magnetization measurements therefore provide an experimental tool to
discriminate between the two possible order parameter structures uncovered in
Ref\cite{denisov09}.

\subsection{Mean field}

We start by a short reviewing of the mean-field properties of the functional
\begin{align}
H  &  =\alpha\left\vert \Psi\right\vert ^{2}-\sum_{i=x,y,z}g\left\vert
D_{i}\Psi\right\vert ^{2}+\gamma\left\vert \sum_{i=x,y,z}D_{i}^{2}\right\vert
^{2}\nonumber\\
&  +\varepsilon_{z}\left\vert D_{z}^{2}\Psi\right\vert ^{2}+\varepsilon
_{x}(\left\vert D_{x}D_{y}\Psi\right\vert ^{2}+\left\vert D_{y}D_{x}%
\Psi\right\vert ^{2})\nonumber\\
&  +\varepsilon(\left\vert D_{x}D_{z}\Psi\right\vert ^{2}+\left\vert
D_{z}D_{x}\Psi\right\vert ^{2})\nonumber\\
&  +\varepsilon(\left\vert D_{y}D_{z}\Psi\right\vert ^{2}+\left\vert
D_{z}D_{y}\Psi\right\vert ^{2})
\end{align}
consistent with the tetragonal symmetry of CeCoIn$_{5}$. The terms
$\varepsilon_{z},$ $\varepsilon_{x}$ and $\varepsilon$ describe nontrivial
(namely different from a simple elliptical) anisotropy \cite{denisov09}. Note
that the cubic symmetry corresponds to $\varepsilon_{x}=\varepsilon$ and
$\varepsilon_{z}=0$. It was shown that two kinds of modulated superconducting
states (scenarios a) and b) mentioned above in the general introduction) are
the most likely to occur when anisotropies are properly taken into account.

\textit{The class a) of solutions }corresponds to order parameters modulated
along the field with characteristic FFLO wave-vector $q_{0}$ and in the $n=0$
Landau level. Following the mean field analysis of Ref\cite{denisov09}, we
write the fluctuation spectrum as
\begin{equation}
E_{n=0}(k_{z})=\tau+\gamma_{a}\left(  \dfrac{2eH}{\hslash}+k_{z}^{2}-q_{0}%
^{2}\right)  ^{2} \label{spectrumclassa}%
\end{equation}
which indicates an instability towards finite modulation along the $z$ axis
(magnetic field). This is similar than Eq.($\ref{spectreLandau}$) except that
the Landau index is fixed $n=0$ and $\gamma_{a}$ is a renormalized parameter
(specific to this class a) of solutions) which depends on $\varepsilon_{z},$
$\varepsilon_{x}$ and $\varepsilon$.

\textit{In the class b) of solutions}, the modulation occurs in the plane
perpendicular to the field and is described by a higher ($n>0$) Landau level.
Following the mean field analysis of Ref\cite{denisov09}, we write the
fluctuation spectrum as
\begin{equation}
E_{n}(k_{z})=\tau+\gamma_{b}(Q_{n}^{2}-q_{0}^{2})^{2}+g_{b}k_{z}^{2}.
\label{spectrumclassb}%
\end{equation}
\textit{ }This spectrum differs from Eq.$\left(  \ref{spectrumclassa}\right)
$ since the kinetic energy term $g_{b}k_{z}^{2}$ favors $k_{z}=0$. Furthermore
a finite Landau index is possible, since the energy is also minimized by
choosing $n$ such as $Q_{n}^{2}-q_{0}^{2}\sim0$. In brief, the lowest energy
fluctuations in the normal state ressemble the superconducting groundstate
which is modulated perpendicularly to the field. Note that here $\gamma_{b}$
and $g_{b}$ are also renormalized parameters which are specific to the class
b) of solutions and depend on $\varepsilon_{z},$ $\varepsilon_{x}$ and
$\varepsilon$ in a complicated manner \cite{denisov09}.

\subsection{Fluctuation magnetization}

Here we evaluate the magnetization induced by the FFLO fluctuations taking
into account the intrinsic anisotropy present in 3D compounds. The FFLO
transition might happen under low or strong field, depending on the
underlaying microscopic mechanism. For instance, in the rare earth magnetic
superconductor ErRh$_{4}$B$_{4}$ a small field is sufficient to polarize the
internal moments, and the FFLO transition is thus expected at low applied
magnetic field \cite{bulaevskii85}. Here we treat the case of the FFLO
transition occuring under strong magnetic field which is relevant for
the\ case of the heavy fermion superconductor CeCoIn$_{5}$. Using a single
Landau level approximation, we demonstrate that the magnetization exhibits
qualitatively distinct behaviors depending on the class of solutions.

\textit{FFLO-like modulation along the field, characterized by a finite
wave-vector }$q_{0}$\textit{ and Landau index }$n=0$\textit{ (scenario a)
discussed in the introduction). }The 3D density of free energy is given by the
integral
\begin{equation}
F=k_{B}T\frac{H}{\Phi_{0}}\int\frac{dk_{z}}{2\pi}\ln\frac{E_{n=0}(k_{z})}{\pi
k_{B}T}%
\end{equation}
where the energy $E_{n=0}(k_{z})$ is given by Eq.$\left(  \ref{spectrumclassa}%
\right)  $. Hence the most divergent part of the orbital magnetization (per
unit volume) $M=-\partial F/\partial H$ is given by
\begin{align}
M  &  =-\frac{2k_{B}T}{\Phi_{0}^{2}}\times\nonumber\\
&  \times\left(  \int_{-\infty}^{\infty}dk_{z}\frac{\gamma_{a}\left(
2eH/\hslash+k_{z}^{2}-q_{0}^{2}\right)  }{\tau+\gamma_{a}\left(
2eH/\hslash+k_{z}^{2}-q_{0}^{2}\right)  ^{2}}\right)  H,
\end{align}
where $\Phi_{0}=h/(2e)$. Since the numerator of the integrand cancels and
changes sign as a function of $k_{z}$, one expects a strong suppression of the
fluctuation magnetization compared to the uniform BCS case wherein such a
cancellation does not occur. Indeed the magnetization $M$ diverges
logarithmically at the FFLO transition which is less divergent than the
$\tau^{-1/2}$ law predicted in the standard BCS case. Therefore the presence
of a genuine FFLO state should be detected as a suppression of the fluctuation
diamagnetism observed near the BCS transition. In comparison with the 2D case,
the oscillations between paramagnetism and diamagnetism predicted in the
previous section are blurred out by the dispersion over the momentum $k_{z}$
along the field.

\begin{figure}[ptb]
\begin{center}
\includegraphics[width=3.5in,angle=0]{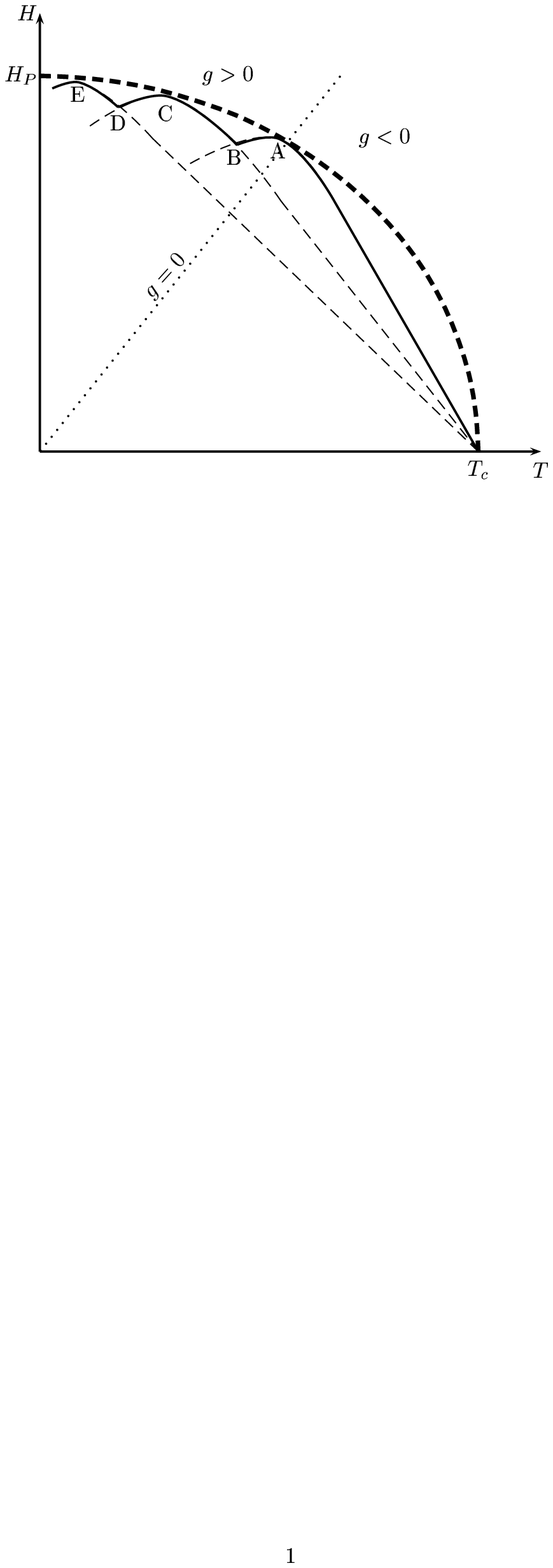}
\end{center}
\caption{Schematic field-temperature $\left(  H,T\right)  $ phase diagram
showing the cascade of Landau levels. The thick dashed curve represents the
critical field in the absence of orbital effect. In presence of orbital
effect, the critical field is reduced and described by the thin dashed curves
which corresponds respectively to the $n=0,$ $n=1$ and $n=2$ Landau levels.
The solid line represents schematically the expected transition line between
the normal and the superconducting states.\ The fluctuations are diamagnetic
between zero field ($H=0,T_{c}$) and $B$, paramagnetic near the arc $AB$, then
again diamagnetic near the arc $BC$, etc...The Landau levels are degenerate at
points $B$,$D$, etc...\ This schematic picture is relevant for both 2D FFLO
superconductors and for 3D ones where the Landau level modulation is realized
(scenario b) evoked in the introduction).}%
\label{FIG_phase_diagram_anisotropic_functional}%
\end{figure}

\textit{Landau level modulation perpendicular to the field (scenario b)
discussed in the introduction). }The dispersion of the fluctuations
Eq.$\left(  \ref{spectrumclassb}\right)  $ now favors the absence of
modulation along the $z$ axis in contrast to the spectrum
Eq.(\ref{spectrumclassa}). Upon increasing the parameter $q_{0}^{2}$, the
lowest energy Landau level is successively $n=0$, then $n=1$ etc... Near the
BCS transition ($g>0$), the fluctuations induce diamagnetism and a lowering of
the critical field $H_{c2}(T)$ below the purely paramagnetic critical field
$H_{P}(T)$ at the same temperature
(Fig.\ref{FIG_phase_diagram_anisotropic_functional}). When the $n$-th Landau
level is realized and when all the other Landau levels are distant in energy,
one can single out the contribution of this main level to the density of free
energy
\begin{equation}
F=k_{B}T\frac{H}{\Phi_{0}}\int_{-\infty}^{\infty}\frac{dk_{z}}{2\pi}\ln
\frac{\tau+\gamma_{b}(Q_{n}^{2}-q_{0}^{2})^{2}+g_{b}k_{z}^{2}}{\pi k_{B}T}.
\end{equation}
Writing the orbital magnetization as
\begin{align}
M_{n}  &  =-\frac{(4n+2)k_{B}T}{\Phi_{0}^{2}}\times\nonumber\\
&  \times\left(  \int_{-\infty}^{\infty}dk_{z}\frac{\gamma_{b}(Q_{n}^{2}%
-q_{0}^{2})}{\tau+\gamma_{b}(Q_{n}^{2}-q_{0}^{2})^{2}+g_{b}k_{z}^{2}}\right)
H_{\perp},
\end{align}
shows that the 2D oscillations are no longer suppressed by the integration
over $k_{z}$ since the numerator is independent of $k_{z}$. Calculating the
integral shows that the magnetization diverges as
\begin{equation}
M_{n}=-\frac{(4n+2)k_{B}T}{\Phi_{0}^{2}}\frac{\gamma_{b}}{\left(
ag_{b}\right)  ^{1/2}}\frac{Q_{n}^{2}-q_{0}^{2}}{\left(  T-T_{cn}\right)
^{1/2}}H_{\perp}%
\end{equation}
with the same power law than in the BCS transition of 3D superconductors
\cite{b.larkin_varlamov}. Unlike the BCS case, this fluctuation magnetization
changes sign being diamagnetic when $Q_{n}^{2}>q_{0}^{2}$ (arcs $BC$, $DE$ in
Fig.\ref{FIG_phase_diagram_anisotropic_functional}) and paramagnetic when
$Q_{n}^{2}<q_{0}^{2}$ (arcs $AB$, $CD$).

In brief, the magnetization is sizeable and oscillates between para- and
diamagnetism when the superconducting order parameter is modulated
perpendicularly to the field, whereas it is strongly suppressed when the order
parameter is modulated along the field. Therefore magnetization measurements
may serve as a test to discriminate between FFLO and Landau level like
modulations in 3D anisotropic superconductors.

\ \ 

\bigskip

\section{Conclusion.\label{PAR_discussion_crossovers}}

We investigated the conductivity and the orbital magnetization associated with
superconducting fluctuations above the FFLO critical temperature or field.
Both in 2D and 3D models, we shown that these properties differ considerably
than their counterparts at the vicinity of a standard BCS transition towards
an homogeneous superconducting state, thereby providing an experimental tool
to detect the inhomogeneous state. First, the paraconductivity of thin
superconducting films exhibits a strong anisotropy when measured parallel or
perpendicular to the FFLO modulation. Second, the orbital magnetization
oscillates between diamagnetic and paramagnetic behaviors at high fields, and
is strongly suppressed at low fields, whereas the uniform BCS state always
induces diamagnetic fluctuations above $T_{c}$. We suggest performing
magnetization and conductance measurements along the FFLO transition line in
compounds where the FFLO state has been recently reported. In 2D organic
superconductors \cite{uji.2006,lortz.2007}, the magnetization oscillations
should be even more pronounced than in the 3D magnetic superconductors
(ErRh$_{4}$B$_{4}$ , see \cite{bulaevskii85}) or in the case of the
anisotropic 3D heavy fermion superconductor CeCoIn$_{5}$%
\cite{radovan03,bianchi02,bianchi03}. It was recently shown that CeCoIn$_{5}$
has quasi-2D Fermi-surface sheets coexisting with a 3D anisotropic Fermi
surface \cite{ko}. Nevertheless due to strong hybridization, superconductivity
in CeCoIn$_{5}$ is likely to be described by a single order parameter. The
complex structure of the Fermi surface should only modify the expressions of
the coefficients in the MGL functional as functions of the microscopic
parameters of CeCoIn$_{5}$. Moreover, the fact that the Land\'{e} factor is
anisotropic leads to different Maki parameters depending on the field
orientation and may also shift the position of the tricritical point
\cite{shimahara_matsuda}. Nevertheless the magnetization
oscillations/suppression predicted here are generic of the presence of FFLO
phase, and should pertain independently of the microscopic characteristics of
CeCoIn$_{5}$.

Finally, in the 3D case, we find that the absence of such oscillations reveals
a FFLO state modulated along the field whereas presence of oscillations should
be associated with a multiquanta Landau modulation perpendicular to the field.

\bigskip

\begin{acknowledgments}
The authors thank Y. Matsuda, D. Denisov, A. Levchenko and M. Houzet for
useful discussions. This work was supported by ANR Extreme Conditions
Correlated Electrons (ANR-06-BLAN-0220).
\end{acknowledgments}

\bigskip

\section{APPENDIX\bigskip}

In this appendix, we address the validity of the Gaussian approximation used
in this paper. When the temperature is sufficiently close to the critical one,
interactions between the fluctuation modes become so strong that the Gaussian
approximation breaks down. In order to quantify the range of temperature where
this breakdown occurs, we derive the Ginzburg-Levanyuk criterion for the FFLO
transition \cite{b.landau.V_e,levanyuk.1959,ginzburg.1960}.

\subsection{Ginzburg-Levanyuk criterion for the FFLO transition}

The full isotropic MGL functional $H\left[  \Psi\right]  =H_{G}\left[
\Psi\right]  +H_{int}[\Psi]$ contains a quadratic part
\begin{align}
H_{G}\left[  \Psi\right]   &  =N_{d}(0)\int d\boldsymbol{r}\left[
\widetilde{\alpha}\left\vert \Psi\right\vert ^{2}-\widetilde{g}\xi
^{2}\left\vert \partial\Psi\right\vert ^{2}+\xi^{4}\left\vert \partial^{2}%
\Psi\right\vert ^{2}\right] \\
&  =N_{d}(0)\sum_{\mathbf{k}}\left(  \widetilde{\tau}+(\mathbf{k}^{2}%
-q_{0}^{2})^{2}\xi^{4}\right)
\end{align}
which describes the free dynamics of the order parameter $\Psi,$ and non
quadratic terms $H_{int}[\Psi]$ which describe
interactions\cite{buzdin_kachkachi(1997),yang_agterberg.2001,houzet2006}. All
the results of this paper are derived within the Gaussian approximation which
consists in using $H_{G}\left[  \Psi\right]  $ as the free energy functional
thereby neglecting completely $H_{int}\left[  \Psi\right]  $. In the spirit of
the original \cite{levanyuk.1959,ginzburg.1960} and
textbook\cite{b.landau.V_e} Ginzburg-Levanyuk criterion, we evaluate the
interaction terms
\begin{equation}
H_{int}\left[  \Psi\right]  =N_{d}(0)\int d\boldsymbol{r}\left[
\frac{\widetilde{g}}{T_{c}^{2}}\left\vert \Psi\right\vert ^{4}+\frac{1}%
{T_{c}^{4}}\left\vert \Psi\right\vert ^{6}\right]  .
\end{equation}
in order to compare them with $H_{G}\left[  \Psi\right]  $.

We have introduced the dimensionless coefficients $\widetilde{\alpha
}=(T-\tilde{T}_{c})/\tilde{T}_{c}$, $\widetilde{\tau}=(T-T_{c})/T_{c}$, and
$\widetilde{g}$ to make apparent the order of magnitude of ech term in the MGL
functional as a function of the the energy scales $T_{c}$ and $E_{F}$. In
particular the dimensionless parameter $\widetilde{g}$ is of order one. We
have also introduced the $d$-dimensional electronic density of states
$N_{d}(0)$ and the superconducting coherence length $\xi=v_{F}/T_{c}$ (we set
$\hslash=1$). Here $q_{0}^{2}=\widetilde{g}/2\xi^{2}$ in analogy with the
transformation performed in Eq. ($\ref{spectrumiso}$). It is a rather
particular property of the MGL functional that the coefficients of the
$\left\vert \Psi\right\vert ^{4}$ and $\left\vert \partial\Psi\right\vert
^{2}$ vanish at the same point $(H,T)$ of the phase diagram (the tricritical
point) \cite{buzdin_kachkachi(1997)}. For this reason and since we are solely
interested in orders of magnitude here, we have denoted the coefficient of the
$\left\vert \Psi\right\vert ^{4}$ term by the same $\widetilde{g}$ used for
the $\left\vert \partial\Psi\right\vert ^{2}$.

For examples of phase transitions, one usually evaluates only the $\left\vert
\Psi\right\vert ^{4}$ terms \cite{b.landau.V_e}. Here the situation of the
FFLO transition is rather specific since on the line $\widetilde{g}=g=0$ of
the $(T,H)$ phase diagram the coefficient of the $\left\vert \Psi\right\vert
^{4}$ term vanishes. Therefore one should evaluate the next interacting term,
$\left\vert \Psi\right\vert ^{6}$, for the regions near this line
$\widetilde{g}=0$. Sufficiently far away from this line $\widetilde{g}=0$ (see
the quantitative criterion below), one may simply evaluate the $\left\vert
\Psi\right\vert ^{4}$ term.

Using Wick theorem to evaluate $H_{int}\left[  \Psi\right]  =H_{int}%
^{(4)}\left[  \Psi\right]  +H_{int}^{(6)}\left[  \Psi\right]  $, we find%
\begin{equation}
H_{int}^{(4)}\left[  \Psi\right]  =\frac{2\widetilde{g}N_{d}(0)}{T_{c}%
^{2}L^{d}}\sum_{\mathbf{k,k}^{\prime}}\left\langle \left\vert \Psi
_{\mathbf{k}}\right\vert ^{2}\right\rangle _{0}\left\langle \left\vert
\Psi_{\mathbf{k}^{\prime}}\right\vert ^{2}\right\rangle _{0}%
\end{equation}
for the $\left\vert \Psi\right\vert ^{4}$ term, and
\begin{equation}
H_{int}^{(6)}\left[  \Psi\right]  =N_{d}(0)\sum_{\mathbf{k}}\left(  \frac
{2}{T_{c}^{2}L^{d}}\sum_{\mathbf{k}^{\prime}}\left\langle \left\vert
\Psi_{\mathbf{k}^{\prime}}\right\vert ^{2}\right\rangle _{0}\right)
^{2}\left\langle \left\vert \Psi_{\mathbf{k}}\right\vert ^{2}\right\rangle
_{0},
\end{equation}
for the $\left\vert \Psi\right\vert ^{6}$ term. Note that in this problem the
form of the free field correlator
\begin{equation}
\left\langle \left\vert \Psi_{\mathbf{k}}\right\vert ^{2}\right\rangle
_{0}=\frac{\pi k_{B}T_{c}/N_{d}(0)}{\widetilde{\tau}+\xi^{4}\left(
k^{2}-q_{0}^{2}\right)  ^{2}}. \label{Eq_corrgene}%
\end{equation}
is rather special due to the proximity of the FFLO transition.

\subsection{Isotropic model}

\textit{Far from the tricritical point, namely when }$\widetilde{g}%
\geqslant(T_{c}/E_{F})^{2(d-1)/(6-d)}$\textit{, }the leading correction to the
Gaussian behavior originates from the $\left\vert \Psi\right\vert ^{4}$
interaction term. Morover the fluctuations propagate with a quadratic
dispersion, and the correlator Eq.($\ref{Eq_corrgene}$) can be approximated
as
\begin{equation}
\left\langle \left\vert \Psi_{\mathbf{k}}\right\vert ^{2}\right\rangle
_{0}=\frac{\pi k_{B}T_{c}/N_{d}(0)}{\widetilde{\tau}+4q_{0}^{2}\xi^{4}\left(
k-q_{0}\right)  ^{2}}%
\end{equation}
when evaluating the sum%
\begin{equation}
\frac{\widetilde{g}}{T_{c}^{2}L^{d}}\sum_{\mathbf{k}^{\prime}}\left\langle
\left\vert \Psi_{\mathbf{k}^{\prime}}\right\vert ^{2}\right\rangle _{0}%
=\frac{\widetilde{g}^{d/2}\widetilde{\tau}^{-1/2}}{T_{c}N_{d}(0)\xi^{d}}.
\end{equation}
Using $T_{c}N_{d}(0)\xi^{d}=(E_{F}/T_{c})^{d-1}$, we find that the $\left\vert
\Psi\right\vert ^{4}$ interaction terms are negligible in comparison to the
Gaussian ones when the condition (Ginzburg-Levanyuk criterion)%
\begin{equation}
\widetilde{\tau}\gg\widetilde{g}^{d/3}\left(  \frac{T_{c}}{E_{F}}\right)
^{2(d-1)/3} \label{EQ_GL1}%
\end{equation}
is fullfilled. The critical region width is larger than in the standard BCS
case ($\tau\gg\left(  T_{c}/E_{F}\right)  ^{4}$ for $d=3$ and $\widetilde
{\tau}\gg T_{c}/E_{F}$ for $d=2$) but it remains extremelly thin.

\textit{Near the tricritical point, when }$\widetilde{g}\leqslant(T_{c}%
/E_{F})^{2(d-1)/(6-d)}$\textit{, }the $\left\vert \Psi\right\vert ^{6}$
interaction becomes stronger than the $\left\vert \Psi\right\vert ^{4}$ one
since this latter contribution is suppressed by the extremelly small prefactor
$\widetilde{g}$. In particular, along the $\widetilde{g}=g=0$ line in the
($H,T$) diagram, the $\left\vert \mathbf{\nabla}\Psi\right\vert ^{2}$ and
$\left\vert \Psi\right\vert ^{4}$ terms are totally absent from the functional
\cite{buzdin_kachkachi(1997)}. Therefore one should compute the mean value
$\left\langle \left\vert \Psi_{\mathbf{k}}\right\vert ^{2}\right\rangle _{0}$
with a purely quartic momentum dependence. Since
\begin{equation}
\frac{1}{T_{c}^{2}}\frac{1}{N_{d}(0)L^{d}}\sum_{\mathbf{k}^{\prime}%
}\left\langle \left\vert \Psi_{\mathbf{k}^{\prime}}\right\vert ^{2}%
\right\rangle _{0}=\frac{\widetilde{\tau}^{(d-4)/4}}{T_{c}N_{d}(0)\xi^{d}},
\end{equation}
the condition to neglect the $\left\vert \Psi\right\vert ^{6}$ interaction
between the fluctuation modes is thus
\begin{equation}
\widetilde{\tau}\gg\left(  \frac{T_{c}}{E_{F}}\right)  ^{4(d-1)/(6-d)}%
\end{equation}
The critical fluctuations are present in a larger region of the phase diagram
than for BCS superconductivity \cite{b.larkin_varlamov}. During the completion
of this work, we became aware of Ref.\cite{zyuzin09} where the
Ginzburg-Levanyuk criterion is derived by evaluting exclusively the
$\left\vert \Psi\right\vert ^{4}$ interaction term. We therefore obtain the
same Ginzburg-Levanyuk criterion as in Ref.\cite{zyuzin09} for the large
$\widetilde{g}$ regime whereas our criteria differ when approaching the
tricritical point. In spite of this discrepancy, both procedures lead to the
same practical conclusion that the critical region remains extremelly thin and
inaccessible for experimental observations, because of the smallness of the
ratio $T_{c}/E_{F}\sim(10^{-2}-10^{-3})$.

\subsection{Anisotropic model.}

We now derive the Ginzburg-Levanyuk criterion in the case of anistropic FFLO
superconductors. The large $g$ regime is modified in comparison to the
isotropic case, since there the low energy fluctuations are located around few
isolated points instead being spread over a large shell of radius $q_{0}$.

\textit{Far from the tricritical point, namely when }$\widetilde{g}%
\geqslant(T_{c}/E_{F})^{2(d-1)/(6-d)}$\textit{, }the leading correction to the
Gaussian behavior originates from the $\left\vert \Psi\right\vert ^{4}$
interaction term. Moreover the fluctuations propagate with a quadratic
dispersion, and the correlator Eq.($\ref{Eq_corrgene}$) can be approximated
as
\begin{equation}
\left\langle \left\vert \Psi_{\mathbf{k}}\right\vert ^{2}\right\rangle
_{0}=\frac{\pi k_{B}T_{c}/N_{d}(0)}{\widetilde{\tau}+\widetilde{g}k^{2}\xi
^{2}}.
\end{equation}
Evaluating the sum%
\begin{equation}
\frac{\widetilde{g}}{T_{c}^{2}L^{d}}\sum_{\mathbf{k}^{\prime}}\left\langle
\left\vert \Psi_{\mathbf{k}^{\prime}}\right\vert ^{2}\right\rangle
_{0}=\left(  \frac{\tau}{\widetilde{g}}\right)  ^{^{(d-2)/2}}\frac{1}%
{T_{c}N_{d}(0)\xi^{d}}%
\end{equation}
and using $T_{c}N_{d}(0)\xi^{d}=(E_{F}/T_{c})^{d-1}$, we find that the
$\left\vert \Psi\right\vert ^{4}$ interaction terms are negligible in
comparison to the Gaussian ones when the condition (Ginzburg-Levanyuk
criterion)%
\begin{equation}
\widetilde{\tau}\gg\widetilde{g}^{(2-d)/(4-d)}\left(  \frac{T_{c}}{E_{F}%
}\right)  ^{2(d-1)/(4-d)}%
\end{equation}
is fullfilled. This Ginzburg-Levanyuk criterion is similar (same power of
$T_{c}/E_{F}$) than the one encountered in the standard BCS case.

\textit{Near the tricritical point, when }$\widetilde{g}\leqslant(T_{c}%
/E_{F})^{2(d-1)/(6-d)}$\textit{, }the lowest energy fluctuations are located
around the origin of the reciprocal space and have a quartic dispersion like
in the isotropic model studied above. The Ginzburg-Levanyuk criterion is thus
again
\begin{equation}
\widetilde{\tau}\gg\left(  \frac{T_{c}}{E_{F}}\right)  ^{4(d-1)/(6-d)}%
\end{equation}
near the tricritical point.\bigskip

\subsection{Conclusion}

We have obtained that the size of the critical region in FFLO superconductors
is more extended than in the usual uniform superconductor case. Nevertheless,
for superconducting compounds, the ratio $T_{c}/E_{F}\sim(10^{-2}-10^{-3})$ is
small and the critical region thus remains hardly accessible for experimental
observations, thereby supporting the Gaussian analysis performed in this
paper. This fact makes very difficult the observation of the phenomena
(first-order transition) predicted by renormalization group studies
\cite{brazovskii.1975,dalidovich_yang.2004}.


\begin{thebibliography}{99}
\bibitem {footnoteLandau}This useful property relies on the isotropy of the
model treated in Sec. II. In the anisotropic case (Sec. III) $\gamma
\mathbf{D}^{4}$ is replaced by a more complicated operator $\gamma
_{ij}\mathbf{D}_{i}^{2}\mathbf{D}_{j}^{2}$ whose eigenmodes differ from the
ones of $\mathbf{D}^{2}$ and are not known exactly.
\end{thebibliography}
\end{document}